\pdfoutput=1 

\documentclass[final,5p,times,twocolumn]{elsarticle}

\usepackage{amssymb}

\usepackage{lineno}

\biboptions{sort&compress }

\usepackage{hyperref}
\usepackage{doi}

\begin{document}

\begin{frontmatter}

\title{SSPALS: a tool for studying positronium}

\author{Adam Deller}
\address{Department of Physics and Astronomy, University College London, \\ Gower Street, London, WC1E 6BT, United Kingdom}
\ead{a.deller@ucl.ac.uk}


\begin{abstract}
Single-shot positron annihilation lifetime spectroscopy (SSPALS) is an extremely useful tool for experiments involving the positronium atom (Ps).  I examine some of the methods that are typically employed to analyze lifetime spectra, and use a Monte-Carlo simulation to explore the advantages and limitations these have in laser spectroscopy experiments, such as resonance-enhanced multiphoton ionization (REMPI) or the production of Rydberg Ps.
\end{abstract}


\end{frontmatter}


\section{Introduction}\label{sec:intro}

Positronium (Ps)~\cite{Ruark1945} is the bound state of an electron and a positron. In vacuum, the components of the particle-antiparticle pair will ultimately annihilate with each other. The mean lifetime against self-annihilation is 125~ps for the $n=1$ singlet spin state ($1^1S_0$, para-Ps)~\cite{Wheeler1946, Al-Ramadhan1994} or 142~ns for the $n=1$ triplet spin states ($1^3S_1$, ortho-Ps)~\cite{Ore1949, Vallery2003}.   Annihilation of p-Ps usually results in two 511~keV gamma-ray photons, whereas o-Ps predominately decays into three with a combined energy of 1.022~MeV~\cite{Rich1981}.  A scintillator coupled to a photomultiplier tube (PMT) can be used to efficiently detect gamma rays with sub-ns timing resolution~\cite{Knoll2010}.  This facilitates precision positron annihilation lifetime spectroscopy (PALS)~\cite{Siegel1980, Peng2005} -- a simple but powerful technique that was instrumental in the discovery of positronium by Deutsch in 1951 \cite{Deutsch1951}. 

The gross atomic structure of Ps~\cite{Ley2002} can be described by the Bohr model for hydrogen but with a reduced mass of $\mu_{\mathrm{Ps}} = m_e /2$; the corresponding energy levels are then given by $E_n = -6.8 / n^{2}$~(eV).  Optical excitation from the ground state can be achieved using a pulsed laser synchronized to a time-bunched Ps source ($\Delta t \lesssim 10$ ns)~\cite{Chu1982, Ziock1990, Cassidy2010, Deller2015a, Wall2015, Aegis2016, Baker2018}. The annihilation and fluorescence decay rates of the excited states range widely~\cite{Alonso2016a, Deller2016a} and laser excitation to these can have a marked effect on the overall lifetime. But to measure a PALS spectrum each annihilation event must be resolvable in the time domain, which is generally not possible with ns Ps sources.  In this case, single-shot positron annihilation lifetime spectroscopy (SSPALS)~\cite{Cassidy2006a} can be implemented instead. Here, the output of a fast gamma-ray detector constitutes the lifetime spectrum. This is a valid approximation if the Ps formation time and the decay time of the detector are both sufficiently short. PbWO$_4$ has a scintillation decay time of $\kappa \sim 10$~ns, which is well suited to resolving o-Ps decay ($\tau = 142$~ns). The Cerenkov radiator PbF$_2$ can be used to improve timing resolution~\cite{Cassidy2007a} but it has a lower light output. For some applications, a slower material with a higher light output, such as LYSO ($\kappa \sim 40$~ns), might be chosen to improve detection efficiency and the signal-to-noise ratio~\cite{Alonso2016}.

Pulsed Ps sources, with time widths of a few ns, can be obtained by implanting time-focused positron beams~\cite{Cassidy2005, Cooper2015} into Ps-converter materials, such as mesoporous SiO$_2$~\cite{Liszkay2008a}.  The interconnected network of pores provides a path to vacuum along which Ps atoms cool via inelastic collisions~\cite{Crivelli2010, Cassidy2010}. The overall efficiency for emission of o-Ps from mesoporous silica is $\epsilon \sim 0.25 $ /e$^{+}$. Other materials are known to be more efficient~\cite[e.g.,][]{MillsJr.1985a}, however, mesoporous silica is often chosen for laser spectroscopy experiments because it performs well at room temperature and the conversion efficiency is generally stable~\cite{Cooper2016}. Moreover, the average kinetic energy of Ps atoms emitted from this material can be relatively low ($40$ -- $100$~meV \cite{Cassidy2010}).  With a mass of $m_\mathrm{Ps} = 2 m_\mathrm{e} \approx 0.0011$~amu, Ps atoms usually move at very high speeds ($|\vec{v}| \approx 1.3 \times 10^5$~ms$^{-1}$ for $KE = 100$~meV).  Consequently, the Doppler width of the $1S\rightarrow2P$ transition is around 0.5~THz for a 400~K distribution. A laser bandwidth $\gtrsim 50$~GHz is therefore required to achieve significant spectral overlap~\cite{Ziock1990, Cassidy2010, Deller2015a}. 
 Alternatively, two-photon Doppler-free excitation schemes can be pursued~\cite{Chu1982, Fee1993}.

There are unique challenges to performing laser spectroscopy with positronium that arise from its low mass and predisposition to annihilate.  On the other hand, annihilation radiation can be harnessed to measure lifetime spectra. In this article,  I describe a simple model for SSPALS spectra and outline the main features and usual methods of analysis, which have been employed in applications ranging from measurement of the ground-state hyperfine interval~\cite{Cassidy2012a} to the discovery of the di-positronium molecule (Ps$_2$)~\cite{Cassidy2007}. Then I present the results of Monte-Carlo (M-C) simulations of Ps distributions resonantly interacting with laser radiation.  I use the simulations to examine how laser excitation of Ps can affect a lifetime spectrum and compare the results with experimental data. 

\section{Single-shot positron annihilation lifetime spectra}\label{sec:sspals}

The time distribution of the gamma radiation generated when a pulsed positron beam impacts a Ps-converter will, in general, contain two main features: (i) a prompt peak associated with the rapid annihilation of e$^{+}$ and p-Ps; and (ii) delayed events that reflect the exponential decay of o-Ps emitted from the converter into vacuum.

\begin{figure}[htp]
    \centering %
    \includegraphics[width=0.4\textwidth]{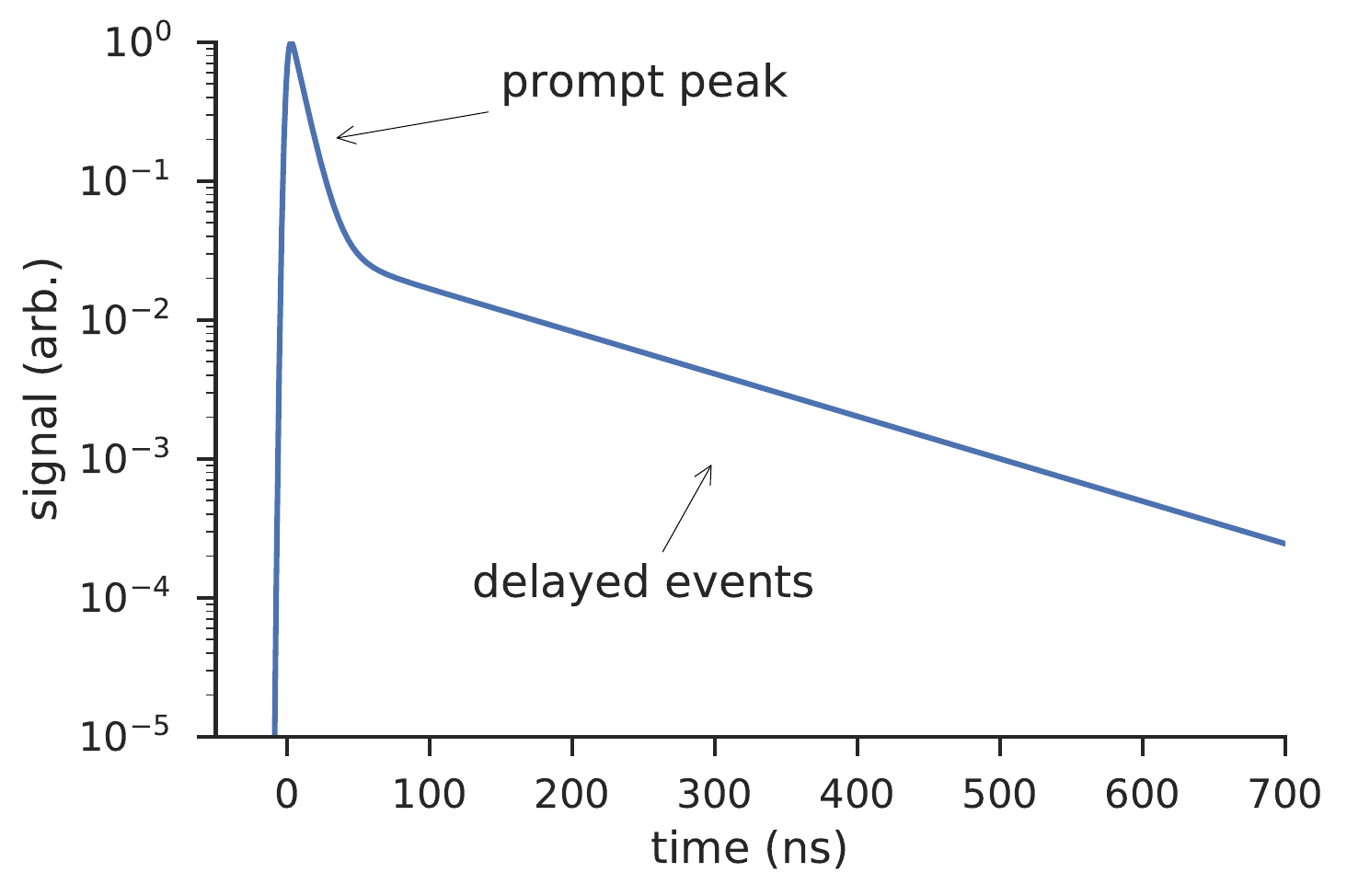}
    \caption{\label{fig:pwo}An SSPALS spectrum modelled using Eq.~\ref{eqn:sspals} ($\epsilon = 0.25$, $\kappa = 9$~ns, and $\sigma = 2$~ns). Scaled for a peak height of 1.}
\end{figure}

An SSPALS spectrum can be analytically modelled using a Gaussian distribution for the positron implantation time (width of $\sigma$), convolved with the $\epsilon$-weighted decay of o-Ps ($\tau = 142$~ns); I use the approximation that unconverted positrons and p-Ps annihilate immediately and that the product of the likelihood of detecting an event and the amplitude of its signal is equal for $3\gamma$ and $2\gamma$ decay.  This function can then be convolved with a model for the detector response (rise time of zero and decay time of $\kappa$).  Altogether, this gives \cite{Deller2013}
\begin{equation}\label{eqn:sspals}
    V(t) = \frac{V_0}{\mathrm{e}^{\frac{t}{\kappa} + \frac{t}{\tau}}} %
            \left(\epsilon \, R_{\kappa, \tau, \sigma}(t)  - %
            \left(1 +  \frac{\tau}{\kappa} \left(\epsilon - 1 \right) \right)  \, R_{\tau, \kappa, \sigma}(t) \right) \;,
\end{equation}
where 
\begin{equation}\label{eqn:Apqr}
    R_{\alpha, \beta, \gamma}(t)  = \mathrm{e}^{\frac{\gamma^{2}}{2 \beta^{2}} + \frac{t}{\alpha}} \, \left(1 + \textrm{erf}\left[\frac{t \beta - \gamma^{2}}{\sqrt{2} \beta \gamma}\right]\right)\;,
\end{equation}
$V_0$ is an arbitrary scaling factor, and $\textrm{erf}$ represents the error function.  Equation~\ref{eqn:sspals} is plotted in Fig.~\ref{fig:pwo} for typical positron pulse and Ps-converter parameters and a detector response appropriate to PbWO$_4$ ($\kappa = 9$~ns).

In principle, Eq.~\ref{eqn:sspals} can be fitted to a measured spectrum to deconvolve its components.  In practice, it is often sufficient to quantify the delayed fraction, $f_d$.  This metric can be used to estimate the Ps conversion efficiency or to observe, e.g., laser-induced changes in the spectra. It is defined as~\cite{Cassidy2011}
\begin{equation}\label{eqn:fd}
f_d = \frac{\int_{B}^{C} V(t) \,dt}{\int_{A}^{C} V(t) \,dt},
\end{equation}
where the time interval $A \rightarrow B$  encompasses the prompt peak and $B \rightarrow C$ contains the delayed events.  Typical choices for these parameters are $A = -10$~ns, $B = 30$~ns, and $C = 700$~ns, where $t = 0$ is given by the positron implantation time. The exact values of $A$ and $C$ are not usually important, so long as the bulk of the spectra is captured. The appropriate value for $B$, however, will depend on the detector response and the process being measured.

\begin{figure}[htp]
    \centering %
    \includegraphics[width=0.4\textwidth]{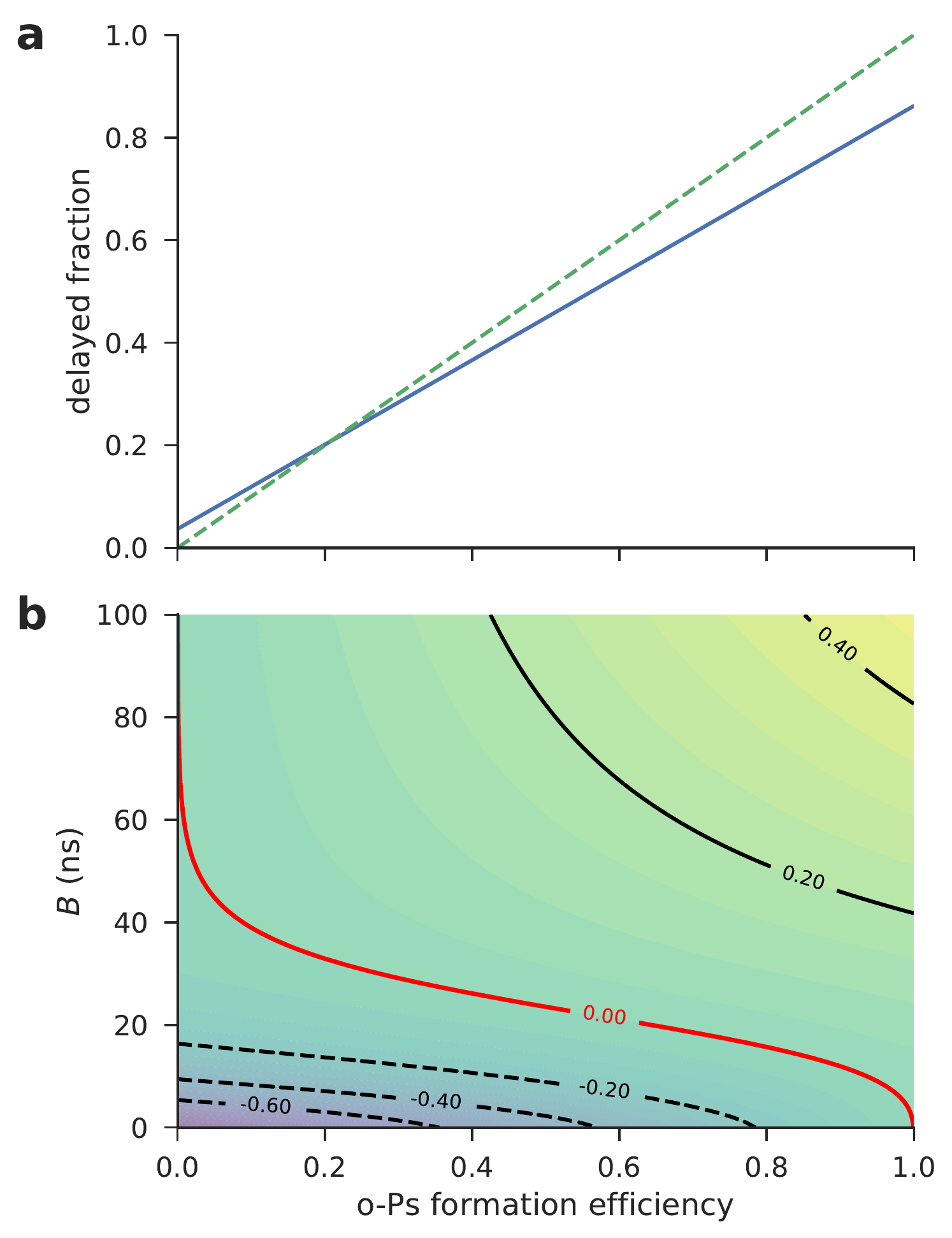}
    \caption{\label{fig:B_val}(a) The delayed fraction, $f_d$, measured for SSPALS spectra modelled using Eq.~\ref{eqn:sspals} ($\tau = 142$~ns, $\kappa = 9$~ns, and $\sigma = 2$~ns) as a function of the Ps conversion efficiency, $\epsilon$. The integration bound $B$ is 30~ns. The dashed line marks $f_d=\epsilon$. (b) Contour plot of the difference between the Ps conversion efficiency and the delayed fraction ($\epsilon - f_d$) as a function of $\epsilon$ and $B$. Negative contours are dashed.}
\end{figure}

In Ref.~\cite{Cassidy2011} the relationship between the Ps conversion efficiency and the measured delayed fraction was examined using an exponential response model for the prompt and delayed components of SSPALS spectra.  Similarly, I calculate $f_d$ for spectra modelled using Eq.~\ref{eqn:sspals} with $\kappa = 9$~ns and $\sigma = 2$~ns. Figure~\ref{fig:B_val}a illustrates that for $B=30$~ns the delayed fraction tracks the Ps conversion efficiency fairly accurately, especially in the region close to $\epsilon = 0.3$.  Figure~\ref{fig:B_val}b demonstrates how the difference between the delayed fraction and the conversion efficiency varies with the choice of $B$, indicating that the broadest range of agreement is found for $B=20$ -- 40~ns.  Note that this depends on the time width of the positron pulse and the time response of the detector \cite{Cassidy2007a, Alonso2016}.

For many applications the absolute value of the o-Ps fraction is not important.  What actually matters is how it changes. This is normally quantified using the parameter \cite{Cassidy2011a}
\begin{equation}\label{eqn:S}
S_\gamma = \frac{f_{bk} - f_d}{f_{bk}},
\end{equation}
where $f_{bk}$ is a background measurement of the delayed fraction (e.g., with the laser tuned off resonance).

\section{Laser spectroscopy}\label{sec:laser}

\subsection{Overview}\label{sec:methods}

\begin{figure}[htp]
    \centering %
    \includegraphics[width=0.3\textwidth]{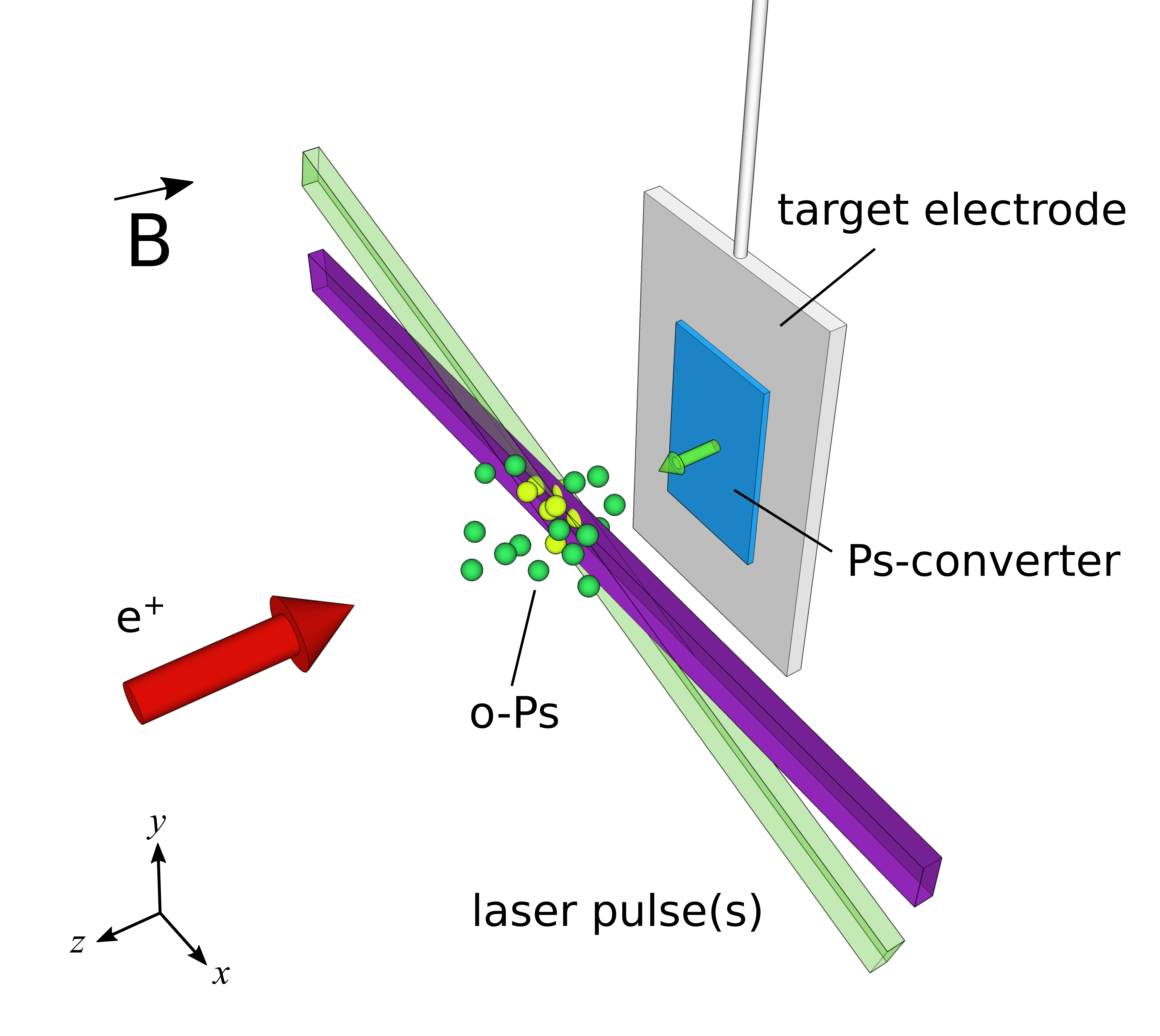}
    \caption{\label{fig:schem} Schematic of a simple Ps-formation and laser-interaction experimental configuration.  A time-focused positron bunch is implanted into a Ps-converter material and lasers intersect the subsequently emitted Ps atoms.}
\end{figure}

A typical arrangement for creating Ps atoms and interrogating them with lasers is shown in Fig.~\ref{fig:schem}.  A magnetic field guides a positron pulse into a Ps-converter material mounted to a planar target electrode. The voltage applied to the electrode is used to tune the positron implantation energy (usually 0.1 - 5~keV). Some of the positrons will form Ps that is emitted to vacuum, where they can be intersected by one or more laser pulses.  A nearby gamma-ray detector measures the SSPALS spectrum.  An additional electrode or a grid (not shown), positioned offset and parallel to the target, can be used to control the electric field in the laser-interaction region~\cite{Wall2015, Alonso2015, Alonso2016a}.

Photoionization of Ps can be performed from the ground state using two-color resonance-enhanced multiphoton ionization (REMPI)~\cite{Cassidy2011a, Deller2015, Aegis2016}. For instance, photoionization via $n=2$ requires an ultraviolet (UV) $\lambda = 243.0$~nm laser for the first transition and an infra-red (IR) or visible $\lambda \leq 729.0$~nm laser to drive the excited state to the continuum. Released positrons can be accelerated by an electric field towards the target and are there likely to annihilate.  The effect upon SSPALS spectra is an annihilation excess that is approximately coincident with the laser pulses, followed by a depletion of the delayed annihilation events associated with ground-state o-Ps. 
The laser-Ps-interaction region is typically chosen to be very close ($< 1$~mm) to the Ps-formation region so that the maximum number of atoms can be addressed before the cloud disperses. Accordingly, photoionization will occur during -- or very shortly after -- the prompt peak.

The 142~ns mean lifetime of ground-state o-Ps imposes a significant restriction on experiments and results in average flight paths of a few cm in vacuum for Ps atoms created in mesoporous silica. However, self-annihilation of Rydberg states ($n \gtrsim 10$) is almost negligible.  These high-$n$ states can be populated using pulsed lasers~\cite{Ziock1990a, Cassidy2012, Wall2015, Jones2014}.   The $1S \rightarrow 2P \rightarrow nS / D$ excitation scheme used in Ref.~\cite{Wall2015} is similar to that for REMPI described above (two-color, two-photon) but the wavelength of the IR laser is in the range of 730 to 770~nm.

The average fluorescence lifetime for Rydberg positronium ranges over $ 3 - 26$~$\mu$s for $n= 10 - 19$~\cite{Deller2016a}. Long flight paths of more than a meter are therefore possible, even for Rydberg Ps atoms with fairly low kinetic energy.  This allows for (almost) background-free detection using, for instance, a micro-channel plate (MCP) removed from the production environment~\cite{Jones2015, Jones2016}.  Nonetheless, SSPALS can also be used for experiments with Rydberg positronium~\cite{Cassidy2012, Wall2015} and it is relatively simple to implement.  The effect that populating Rydberg states has on lifetime spectra depends on several aspects of the experimental arrangement, as discussed in Sec.~\ref{sec:rydberg}.

\subsection{Monte-Carlo simulations}\label{sec:mc}
\subsubsection{positronium distribution}\label{sec:mc_overview}

\begin{figure}[htp]
    \centering %
    \includegraphics[width=0.4\textwidth]{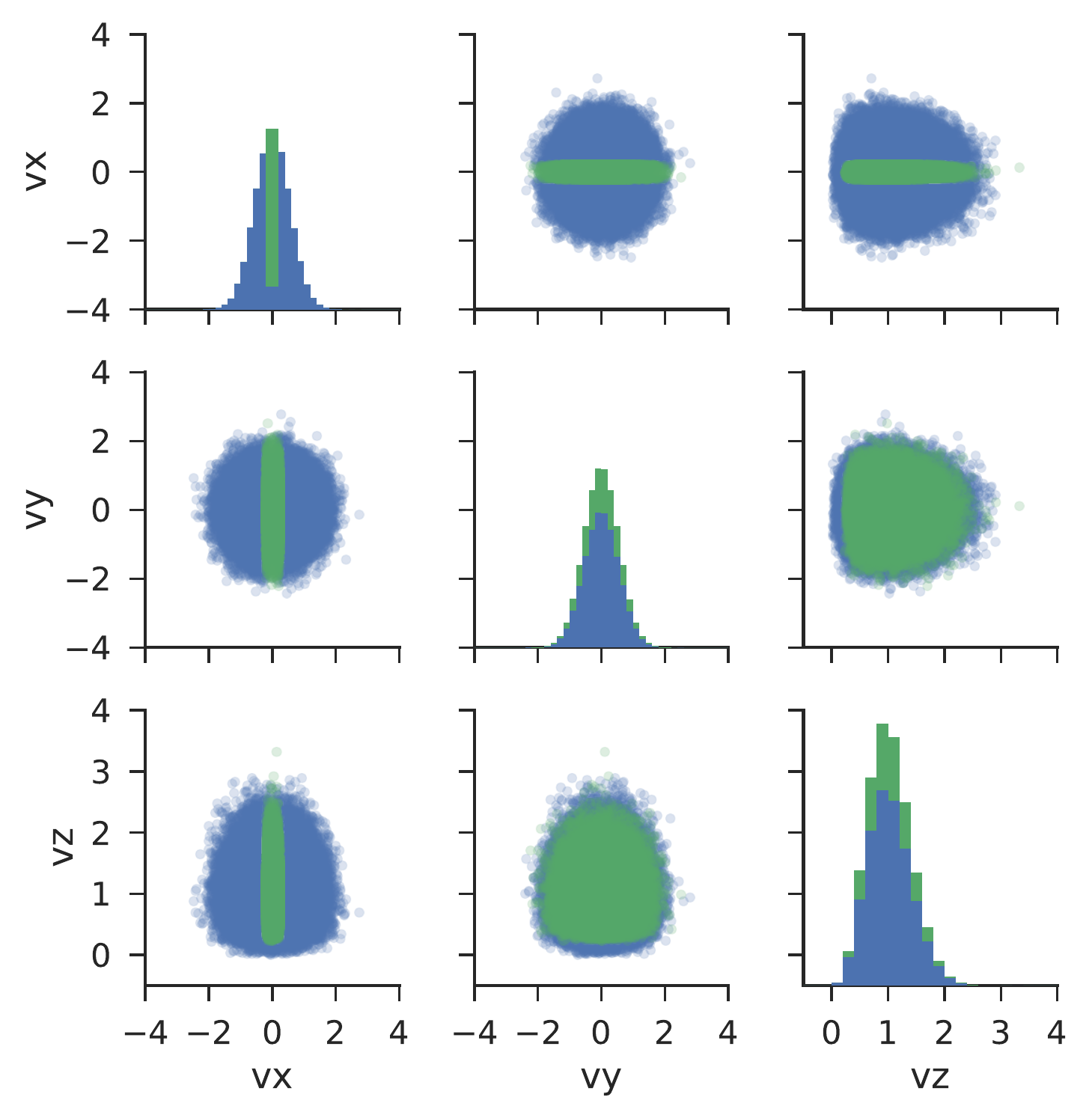}
    \caption{\label{fig:mc_vel} The velocity components (units of $10^5$~m$\,$s$^{-1}$) of a M-C simulated Ps distribution ($T=400$~K).  Atoms excited by a $\lambda = 243.0$~nm, $\Delta\nu = 85$~GHz, $\Delta t = 7$~ns laser pulse are shown in green.}
\end{figure}

A simulation has been made to study the overlap between a Ps source and a pulsed laser field~\cite{adam_deller_2018_1340681}. Monte-Carlo techniques were used to simulate Ps distributions consistent with those created in experimental systems \cite[e.g.,][]{Cooper2015}.  Namely, a million positrons in a 2~mm wide and 5~ns long (FWHM) pulse are converted into o-Ps with an efficiency of $\epsilon = 0.25$ / e$^+$.  The Ps atoms are emitted from a plane surface ($xy$) with a beam Maxwell-Boltzmann velocity distribution.  For $T = 400$~K, the majority of ground-state o-Ps will travel less than 50~mm before undergoing self-annihilation.  At $t=15$~ns a laser pulse ($\Delta t = 7$~ns) passes through the cloud along the $x$-direction, offset by a distance of 0.5~mm from the Ps-converter. The flat rectangular profile of the laser field ($\Delta y = 6$~mm; $\Delta z = 2.5$~mm) mimics that of a pulsed dye laser. The  wavelength is tuned for the $1^3S \rightarrow 2^3P$ transition ($\lambda = 243.0$~nm). The product of the time-integrated laser intensity and spectral overlap experienced by each simulated Ps atom was computed, and those that surpassed a given threshold were deemed to have been excited by the laser, as shown in Fig.~\ref{fig:mc_vel}.

Doppler effects and the laser bandwidth ($\Delta\nu = 85$~GHz) results in laser selection based on the $x$-component of the velocity distribution.  To a lesser extent, the $z$ and $y$ components of an atom's velocity vector will also determine whether or not it can be excited by the laser. This is due to the finite size of the laser profile: some of the Ps atoms pass through the interaction region before or after the laser pulse arrives, or miss it altogether.  The simulation suggests that the laser is able to excite almost $30$~\% of the ground-state Ps atoms.  But if the positron pulse is not radially compressed or not sufficiently bunched, or if the laser is positioned too far from the Ps-converter, this fraction is drastically reduced.

\subsubsection{REMPI}\label{sec:REMPI}

\begin{figure}[htp]
    \centering %
    \includegraphics[width=0.4\textwidth]{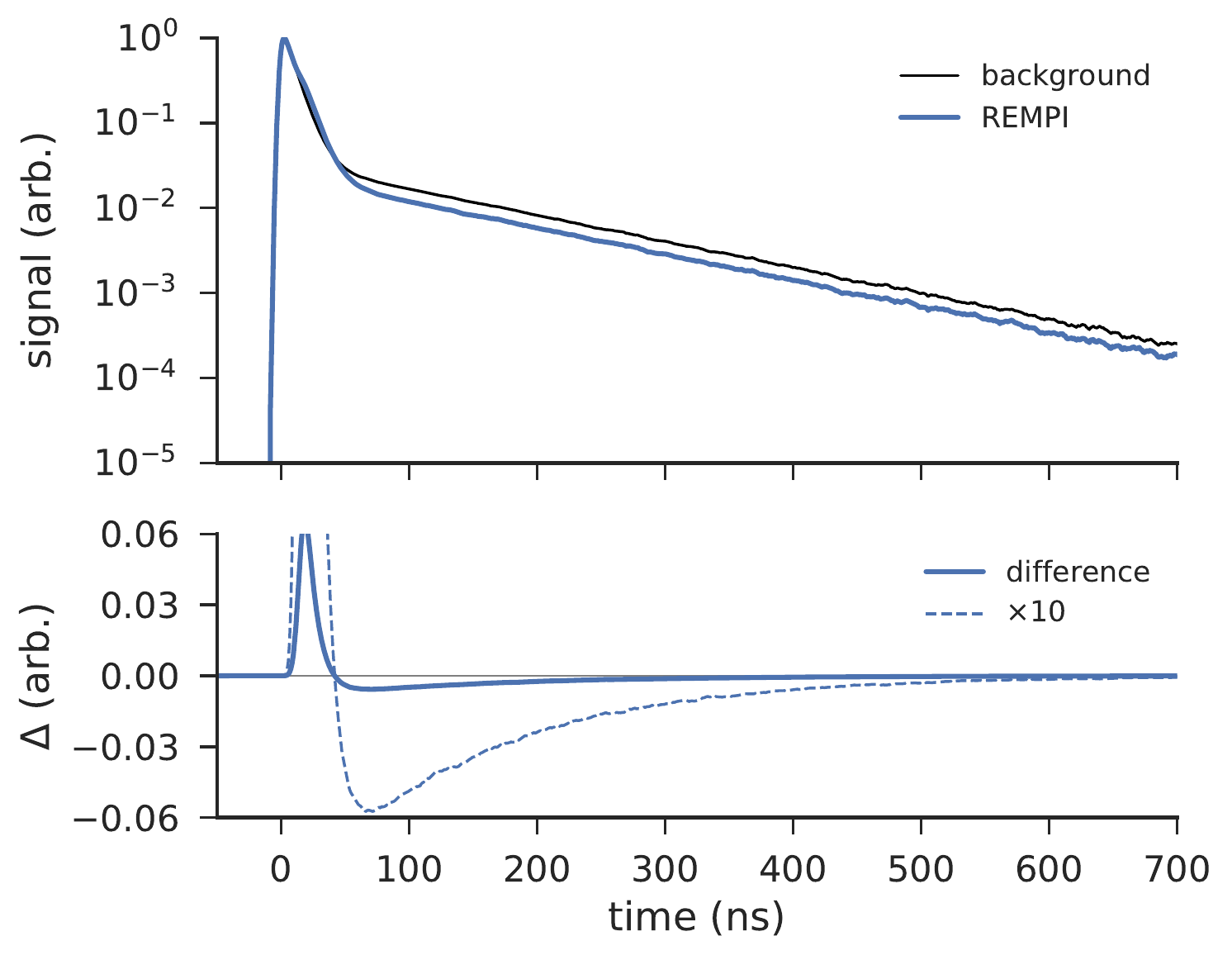}
    \caption{\label{fig:mc_ion} Monte-Carlo simulation of SSPALS spectra ($\epsilon = 0.25$, $\tau = 142$~ns, $\kappa = 9$~ns, $T=400$~K, and $\sigma_t = 2$~ns), including the effect of REMPI via $n=2$.  The UV laser ($\lambda = 243.0$~nm, $\Delta \nu = 85$~GHz) was triggered at $t=15$~ns.  The lower panel shows the background subtracted spectra.}
\end{figure}

Annihilation lifetimes extracted from the M-C simulation were used to generate  SSPALS spectra -- see \ref{appendix-MC} for details. The simulated lifetime spectrum shown in Fig.~\ref{fig:mc_ion} was calculated assuming that those Ps excited by the laser are ionized via REMPI and instantly annihilate.  This causes an increase in the number of annihilation events during the prompt peak, with proportionately fewer after.

In this example, the difference between the background and REMPI simulated SSPALS spectra is a result of $28.0$~\% of the Ps atoms having been photoionized.  These spectra were analyzed using the delayed fraction technique outlined in Sec.~\ref{sec:sspals}, with $A = -10$~ns, $B=35$~ns, and $C=700$~ns  ($t=0$ was found using a constant-fraction-discriminator (CFD) algorithm \cite{Gedcke1967, Deller2016c} that triggers on the rising edge of the prompt peak).  This gives $f_{bk} = 22.5$~\% for the background spectrum, $f_d = 17.3$~\% for the spectrum with the effects of REMPI included, and a value of $S_\gamma = 22.9$~\% for  the difference.

\begin{figure}[htp]
    \centering %
    \includegraphics[width=0.4\textwidth]{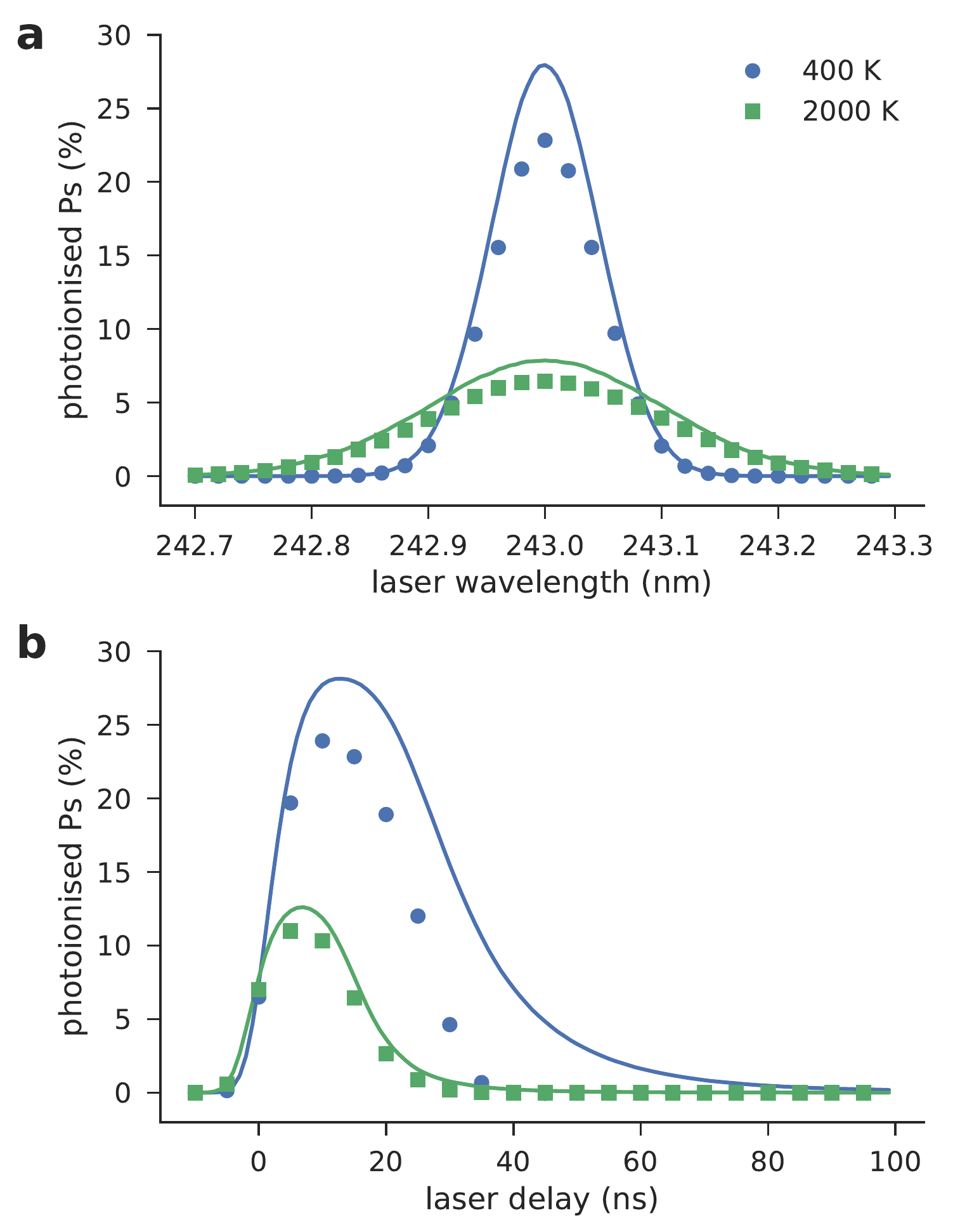}
    \caption{\label{fig:mc_delay_doppler}Monte-Carlo simulation of REMPI via $n=2$, (a) for a range of UV laser wavelengths [t = 15~ns], (b) for different laser delays [$\lambda = 243.0$~nm], and for two different temperature distributions.  The solid lines show the photoionized fraction. The points mark the $S_\gamma$ values obtained from the simulated lifetime spectra by analysing the laser-induced change in $f_d$, using $A=-10$, $B=35$ and $C=700$~ns.}
\end{figure}

Similar lifetime spectra were produced for two different Ps distributions ($T = 400$~K and 2000~K). Figure~\ref{fig:mc_delay_doppler} shows the ionized fraction (lines) compared to values for $S_\gamma$ (points) corresponding to a range of laser wavelengths and delays. 
Figure~\ref{fig:mc_delay_doppler}a illustrates how the temperature of the underlying Ps distribution can be inferred from the measured Doppler width of the Lyman-$\alpha$ transition~\cite{Cassidy2010, Deller2015a, Andersen2015}.  It also demonstrates that, in this case, the $S_\gamma$ parameter is a fairly good estimate for the fractional change in the number of o-Ps atoms. 

However, the laser delay scan plotted in Fig.~\ref{fig:mc_delay_doppler}b shows that $S_\gamma$ does not track the ionized fraction for later laser trigger times. For example, for the $T = 400$~K distribution, $S_\gamma \approx 0$ when the laser was triggered at $t = 40$ ns, even though the ionized fraction was actually $\sim 7$~\%.  This is a consequence of the definition of $f_d$ (Eq.~\ref{eqn:fd}) and the choice of $B$. If a given Ps atom is photoionized after $B$ then the net change to $f_d$ will be zero, as in all likelihood it would have annihilated within the $B \rightarrow C$ time window anyway. This problem can be solved by extending $B$ beyond the laser trigger time, or by choosing integration bounds that track the laser timing. If this type of measurement is performed with a sufficiently well-defined laser position then Ps time-of-flight spectra and longitudinal velocity information can be extracted~\cite{Deller2015}. 

Magnetic quenching \cite{Deutsch1951b, Curry1973} has also been exploited for laser spectroscopy of positronium \cite{Ziock1990, Cassidy2010, Deller2015a, Alonso2015}. 
Laser excitation from the triplet ground state to a magnetically-mixed excited state can lead to spontaneous decay to the short-lived singlet ground state.  For excitation to states that rapidly fluoresce (e.g., $2P$) the overall lifetime is thereby reduced.  The effect had on SSPALS spectra is similar to REMPI, albeit weaker because of competition with decay to the triplet ground state~\cite{Alonso2016a}.

\subsubsection{Rydberg positronium}\label{sec:rydberg}

There are several ways in which exciting Ps atoms to Rydberg levels can affect SSPALS spectra.  A positronium atom is very unlikely to annihilate directly in a high-$n$ state. Accordingly, laser excitation can extend an atom's lifetime by the time it takes it to decay back to the ground state; for $n = 15$, the average fluorescence lifetime is $\sim 10$~$\mu$s~\cite{Deller2016a} (almost two orders of magnitude longer than the ground-state annihilation lifetime).  However, Rydberg states can be ionized by electric fields of just of a few kV~cm$^{-1}$, which are not atypical of the experimental arrangement described in Sec.~\ref{sec:methods}.  Ionization likely leads to the annihilation of the free positron within a few ns. Exciting Ps to Rydberg levels can, therefore, result in lifetime components in SSPALS spectra that are distinctly longer or shorter than the ground-state lifetime, depending on the experimental environment~\cite{Wall2015}.

The Monte-Carlo simulation described in Sec.~\ref{sec:REMPI} was adapted such that, instead of photoionizing Ps, the lasers would drive transitions to states with an average lifetime of 4~$\mu$s. The excitation scheme is a two-step transition via $n=2$. I assume that the second step is 50~\% likely for all of the atoms selected by the laser during the first step.  The effect of Ps atoms colliding with the chamber walls and annihilating was also added to the simulation.  The simulated chamber is a tube with an internal diameter of 36~mm, aligned to the $z$-axis and located a distance of $z = 30$~mm from the Ps-converter.   Typical Ps flight times to the walls range from 0.1 to 1.0~$\mu$s.  A grid electrode with an open area of 90~\% is positioned 9~mm from the target.  This is an approximation to the experimental arrangement described in Ref.~\cite{Wall2015}.~\footnote{Housing the Ps-converter in a fairly small vacuum chamber allows for the SSPALS detector to be positioned very near to the laser-interaction region to maximise the overall detection efficiency.}

\begin{figure}[htp]
    \centering %
    \includegraphics[width=0.4\textwidth]{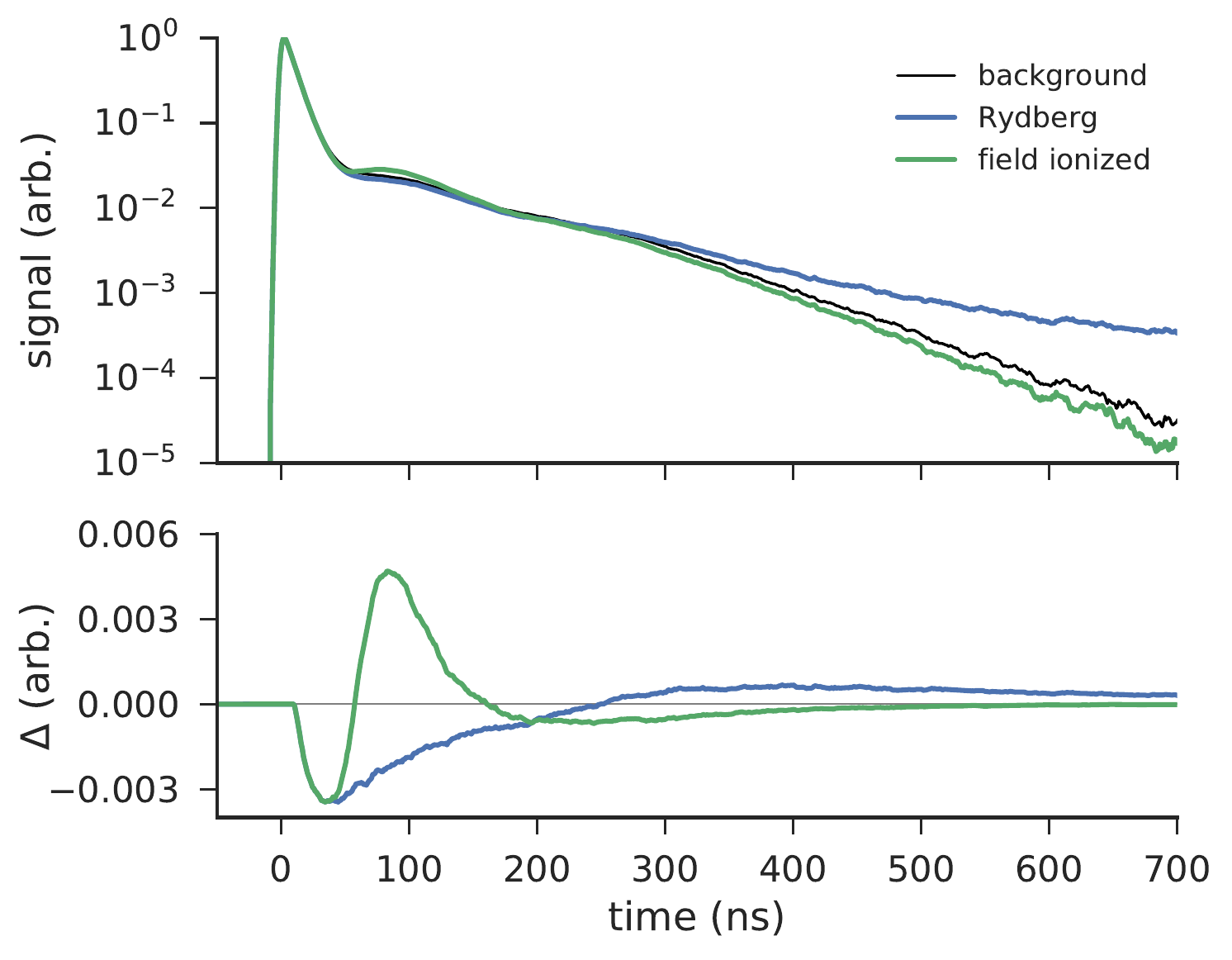}
    \caption{\label{fig:mc_rydberg} Monte-Carlo simulation of SSPALS spectra ($\epsilon = 0.25$, $\tau = 142$~ns, $\kappa = 9$~ns, and $\sigma = 2$~ns), including the effect of populating Rydberg levels that are either long-lived ($\tau = 4$~$\mu$s) or ionized at $z = 9$~mm.  The laser was triggered at $t=15$~ns.  The lower panel shows the background subtracted lifetime spectra.}
\end{figure}

Three simulated SSPALS spectra were generated: (i) with only ground-state o-Ps (background); (ii) with laser-excitation to long-lived Rydberg states; and (iii) where any atoms in Rydberg states that travel a distance of $z = 9$~mm from the Ps converter annihilate there.  The third case is to simulate the effect of a region of high electric field after the grid that ionizes the Rydberg atoms. These spectra are plotted in Fig.~\ref{fig:mc_rydberg}, with the background subtracted spectra displayed in the lower panel.

In all three cases, the inclusion of collisions with the grid and chamber walls adds broad lumps to the spectra at $t \sim 100$ and 200~ns.  Exciting Ps atoms to long-lived Rydberg states (case ii) partially suppresses the early part of the 142~ns lifetime component. An additional long lifetime component becomes apparent at later times~\cite{Cooper2015}.  The delayed events would be evident even if the Rydberg levels had much longer fluorescence lifetimes because of the excited Ps atoms hitting the chamber walls. For  $B = 35$~ns, the delayed fraction analysis gives  $S_\gamma \approx 1.5$~\%.  This apparent insensitivity to Rydberg production can be remedied by setting $B$ to $250$~ns, which yields $S_\gamma = - 38.1$~\%.  The value of $B$ can be tuned to maximize the $S_\gamma$ signal-to-noise ratio~\cite{Alonso2016}. The optimal integration bounds will depend on the Ps velocity distribution, the surroundings, the detector location, the scintillation decay time and background noise levels.

The negative sign of $S_\gamma$ is in contrast to the positive values found for REMPI.  Because $B$ has been set to a later time, $f_d$ is not representative of the number of o-Ps contributing to the spectra (see Sec.~\ref{sec:sspals}).  And although $S_\gamma \sim -40$~\% is an indication of laser excitation events, its value is not a good measure of how many of the atoms have been excited ($\sim 14.0$~\%).  Moreover, the value of $S_\gamma$ will vary significantly with the choice of~$B$. 

A similar analysis of the lifetime spectrum that included field-ionized Rydberg Ps (case iii), using $B = 35$~ns, results in a small signal of $S_\gamma = - 1.3$~\%.  Whereas, for $B = 250$~ns, $S_\gamma \approx 16.1$~\%.  Note that the latter value is positive. The effect of field ionization is broadly similar to REMPI but with the gamma-ray excess delayed by the flight time to the grid ($\sim 100$~ns).  However, it is an accident of the Ps distribution and chamber geometry that the distinct processes of long-lived or field-ionized Rydberg states can be analysed effectively using the exact same integration bounds. In general, the analysis should be optimized for each process separately \cite{Alonso2016}.

\subsection{Experimental data}\label{sec:prl}
An example of the application of SSPALS to laser spectroscopy of Rydberg states of positronium is reported in Ref.~\cite{Wall2015}.  The authors created Ps atoms by implanting 4~keV positrons into mesoporous silica. The emitted Ps distribution was then exposed to UV and IR laser pulses that had been tuned to drive $1S \rightarrow 2P \rightarrow nS/D$ transitions from the ground state to Rydberg levels.  A grid electrode positioned $\sim8$~mm in front of the silica target set the electric field in the laser-interaction region to $|\vec{F}| \approx 0$~kV~cm$^{-1}$. A PbWO$_4$ scintillator optically coupled to a PMT was used to detect annihilation radiation.  The output of the PMT was split between a high gain channel and a low gain channel of a digital oscilloscope, and the data was spliced together in post-processing to simultaneously achieve high-resolution and a wide dynamic range.

\begin{figure}[ht]
    \centering %
    \includegraphics[width=0.4\textwidth]{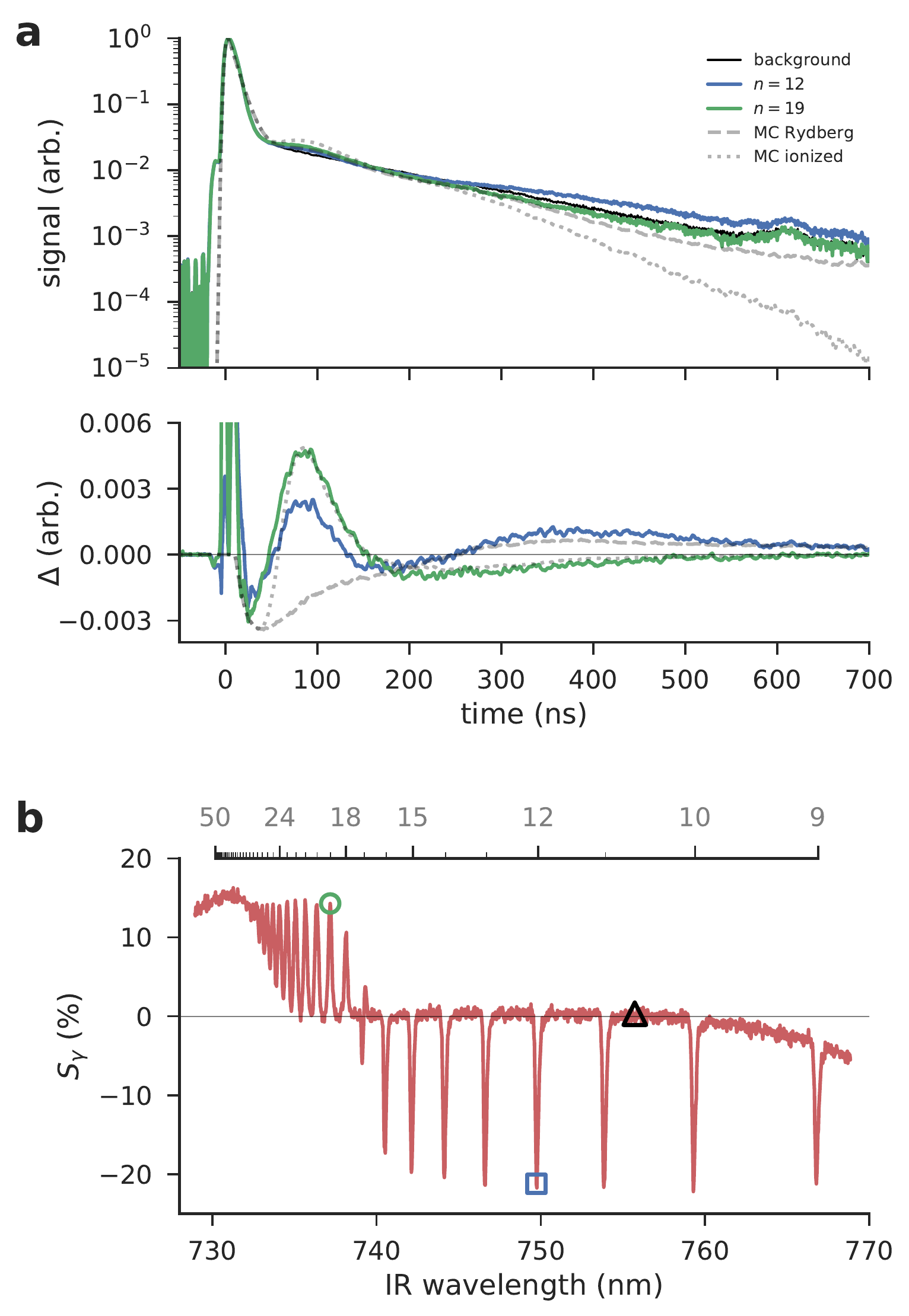}
    \caption{\label{fig:prl_sspals} 
    (a) SSPALS spectra recorded for Rydberg Ps production via two-photon laser excitation ($n=12$ and $n=19$) and with the IR laser off resonance (background).  The lower panel shows the background-subtracted spectra.  The dashed (dotted) lines show the M-C simulated spectra for long-lived (field ionized) Rydberg Ps. 
    (b) $S_\gamma$ values measured for Ps excited to Rydberg levels via $n=2$ in zero applied electric field  ($A=-2$, $B=226$, and $C=597$~ns).  The top axis shows the wavelengths expected to excite each $n$-state.  The open symbols mark the points corresponding to the SSPALS spectra plotted in (a). Experimental data originally published in Ref.~\cite{Wall2015}.}
\end{figure}

Three of the measured SSPALS spectra, corresponding to three different IR laser wavelengths, are shown in Fig.~\ref{fig:prl_sspals}a; Monte-Carlo simulated spectra for long-lived and field-ionised Rydberg Ps (Sec.~\ref{sec:rydberg}) have been overlain for comparison.  All of the spectra have been scaled for a peak height of 1. 

The background-subtracted spectrum recorded when the laser was tuned to resonantly populate $n=19$ is consistent with the simulated spectrum for field ionization of Rydberg atoms at the grid.  This is expected, given that the electric field between the mesh wires of the grid electrode,  $|\vec{F}| \approx 1.5$~kV~cm$^{-1}$, exceeds the classical ionization field for $n=19$ of $1.1$~kV~cm$^{-1}$~\cite{Gallagher2005}.
However, the background subtracted spectrum recorded when the IR laser was tuned to resonantly populate $n=12$ also has features indicative of field ionization at the grid, even though the classical ionization field of $6.9$~kV~cm$^{-1}$ for this state is much larger than any regions of electric field in the experimental apparatus.  This may be due to tunnel ionization, or deflection of the atoms into the mesh caused by Rydberg-Stark acceleration~\cite{Hogan2016, Deller2016d, Jones2017}.  But Fig.~\ref{fig:prl_sspals}a also shows that a significant amount of the $n=12$ atoms survive beyond the grid region and annihilate later.  The delayed features of the $n=12$ lifetime spectrum are similar to those in the M-C simulated Rydberg Ps SSPALS spectrum, and are attributed to collisions with the chamber walls.

Although there is broad agreement between features of the experimental and simulated spectra there are also several important differences.  The simulated spectrum for long-lived Rydberg Ps production (Fig.~\ref{fig:mc_rydberg}) exhibits a faster decay in the signal than was observed in the experiments (Fig.~\ref{fig:prl_sspals}).  This is probably because the time response of PbWO$_4$ contains slow decay components \cite{Belsky1995} that were not included in the simulated spectra.  Also, the simulation does not account for the variation in transition intensity for each final $n$-state.  This explains why the parameters used in the M-C simulation gave a magnitude for the background-subtracted signal that matched the $n=19$ spectrum but underestimated the delayed signal for $n=12$. Other differences can be attributed to the simulation not including the e$^+$ implantation profile nor the Ps cooling dynamics~\cite{Nagashima1995, Takada2000}, which are known to contribute to delayed emission and an epithermal distribution~\cite{Cassidy2010, Cassidy2010a, Deller2015a}.  The velocity distribution will also be affected by the confinement energy of the Ps atoms inside of the pores ~\cite{Mariazzi2008, Crivelli2010}. Furthermore, the detector's solid-angle view of each decay event could affect the lifetime spectra, however, for the geometry described in this work this is not expected to be very significant.

Figure~\ref{fig:prl_sspals}b shows the $S_\gamma$ values taken from the experimental lifetime spectra as the IR laser was scanned over the wavelength range needed to populate $n=9 - \infty$.  The integration bounds used to measure $f_d$ were $A=-2$~ns, $B=226$~ns and $C=597$~ns~. The sharp peaks are correlated with the IR wavelengths expected to excite Ps from $n=2$ to $9 - 28$.  For $n<17$ the peaks are negative, indicating population of Rydberg levels that generally live longer than ground-state atoms.  However, for $n>17$ the peaks are positive, implying average lifetimes that are shorter than those of the ground-state atoms. This is attributed to field ionization of the Rydberg Ps near the grid electrode~\cite{Wall2015}. The broad bandwidth of the UV laser ($\sim 85$~GHz) 
restricted the resolvable states to $n \lesssim 28$.

\section{Concluding remarks}\label{sec:summary}

My objective with this article was to employ a Monte-Carlo simulation to investigate SSPALS as a tool for atomic physics experiments with positronium, in support of reported experimental works~\cite[e.g.,][]{Cassidy2010, Deller2015a, Wall2015, Aegis2016, Alonso2017a} (for a recent review see Ref.~\cite{Cassidy2018}).  The power of SSPALS in these applications is its simplicity and that it can be adapted to several different types of experiment.  Delayed-fraction analysis has proven a robust method for quantifying lifetime spectra, and it's well-suited to measuring various laser-induced effects.  However, more sophisticated analysis methods can extract more information~\cite{Deller2015} and could conceivably improve the signal-to-noise ratio of spectroscopic measurements.  A better understanding of how and why lifetime spectra are affected by the processes discussed in this work could lead to superior techniques for analysis or could be exploited in optimising the designs of new experiments.

\section*{Acknowledgements}
I wish to thank to David Cassidy for many helpful discussions regarding this work. Funding was provided by the EPSRC (Grant No. EP/K028774/1). The Monte-Carlo simulations~\cite{adam_deller_2018_1340681} were written using IPython~\cite{Perez2007}.

\appendix

\section{\label{appendix-MC}Simulating SSPALS spectra}
In Sec.~\ref{sec:mc}, I used a Monte-Carlo simulation to explore the effect that laser excitation of Ps has on SSPALS spectra~\cite{adam_deller_2018_1340681}.  The objective was to find the point in time and space that each positron in a million-strong distribution ultimately annihilates.  The simulation was initialised by generating the time and $xy$ position that the positrons arrive at a Ps-converter, using random sampling of normal distributions ($\sigma_t = 2$~ns and $\sigma_{x} = \sigma_{y} = 1$~mm).  Next, each positron either annihilates immediately or is converted to o-Ps.  This is decided by comparing a random number between 0 and 1 to the expected conversion efficiency of $\epsilon = 0.25$.  Each positron that had been successfully converted into a Ps atom was then assigned its lifetime using an exponential probability distribution with a decay rate of $1 / 142$~ns.  The \emph{``time of death''} was found for all of the o-Ps and unconverted positrons, and these times were collected into a histogram of 1~ns bins.  This histogram was then convolved with the time-response of PbWO$_4$ ($\kappa = 9$~ns) to produce an SSPALS lifetime spectrum that is consistent with Eq.~\ref{eqn:sspals}.  

To evaluate the overlap of the Ps distribution with a laser field the motion of the Ps atoms was added to the simulation.  The atoms were assumed to be emitted from the converter with a Maxwell-Boltzmann beam distribution.  No forces act on them and they travel in straight lines.  Thus, whether or not an atom traversed the excitation region when the laser was triggered could be deduced from its initial position and velocity.~\footnote{Because step-by-step integration isn't necessary a million particle trajectories can be calculated using an ordinary desktop computer in just a few seconds.} Laser-excitation events were assigned using a threshold cut-off for the product of the laser fluence experienced by each atom and the frequency overlap with the laser, including the Doppler shift of the Lyman-$\alpha$ transition arising from the velocity component $v_x$. Most of the atoms in the distribution were insensitive to the exact level of the threshold, which was set assuming the laser intensity was well above saturation.  The lifetimes of the laser-excited atoms were adjusted in accordance with the process being modelled.  For two-step excitation to long-lived Rydberg states, the average lifetime was extended to $4$~$\mu$s.  The time at which each trajectory would intersect the chamber wall was also computed.  10~\% of the atoms that reach the plane of the grid electrode annihilate there and the rest pass through it.  All of the atoms that live long enough to hit the chamber wall were assumed to annihilate with it. 

\begin{figure}[htp]
    \centering %
    \includegraphics[width=0.4\textwidth]{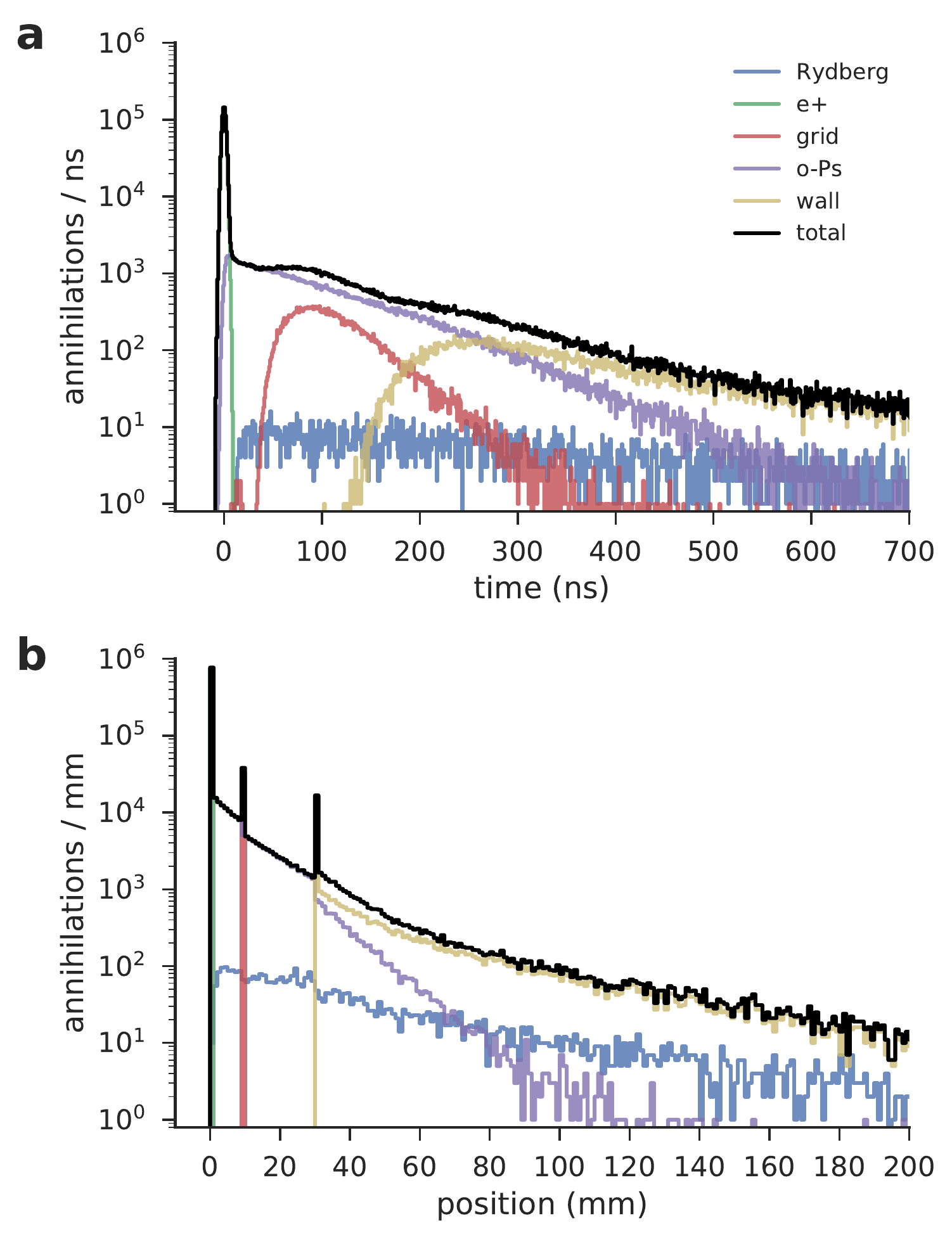}
    \caption{\label{fig:mc_ryd_parts} Histograms of (a) time and (b) $z$-position of a million annihilation events for Monte-Carlo simulated positrons converted to Ps ($T = 400$~K) and excited to Rydberg levels.}
\end{figure}

Histograms of the time and position of annihilation for a simulated Ps distribution are shown in Fig.~\ref{fig:mc_ryd_parts}.  These have been superimposed with histograms that correspond to each of the possible routes to annihilation, namely: e$^+$ (or p-Ps) at the Ps converter, ground-state o-Ps decay in vacuum, Rydberg Ps decay in vacuum, or by collision with the chamber wall or grid.

\bibliographystyle{elsarticle-num}
\bibliography{lib}

\end{document}